%% file: main.tex
\documentclass[journal,preprint]{vgtc}         

\ifpdf
  \pdfoutput=1\relax                   
  \usepackage{graphicx}                
  \DeclareGraphicsExtensions{.pdf,.png,.jpg,.jpeg} 
\else
  \usepackage{graphicx}                
  \DeclareGraphicsExtensions{.eps}     
\fi%

\graphicspath{{figures/}{pictures/}{images/}{./}} 

\usepackage{microtype}                 
\PassOptionsToPackage{warn}{textcomp}  
\usepackage{textcomp}                  
\usepackage{mathptmx}                  
\usepackage{times}                     
\usepackage{cite}                      
\usepackage{tabu}                      
\usepackage{booktabs} 
\usepackage{amsmath}
\usepackage{amsfonts}
\usepackage{multirow}
\usepackage{makecell}
\usepackage{svg}
\usepackage{paralist}

\usepackage{xcolor}
\usepackage{balance}
\usepackage{float}

\definecolor{navy}{RGB}{52, 85, 139}

\ieeedoi{10.1109/TVCG.2021.3114863}

\onlineid{1103}

\vgtccategory{Research}

\title{\techNameNoSpace: A Knowledge Graph-Based Approach for Visualization Recommendation}

\author{Haotian Li, Yong Wang, Songheng Zhang, Yangqiu Song and Huamin Qu}
\authorfooter{
\item
H. Li is with the Hong Kong University of Science and Technology and Singapore Management University. A part of this work was done when he was a visiting student supervised by Yong Wang at Singapore Management University. E-mail: haotian.li@connect.ust.hk
\item
Y. Wang and S. Zhang are with Singapore Management University. Yong Wang is the corresponding author. E-mail: yongwang@smu.edu.sg and shzhang.2021@phdcs.smu.edu.sg
\item
Y. Song and H. Qu are with the Hong Kong University of Science and Technology. E-mail: \{yqsong, huamin\}@cse.ust.hk

}

\shortauthortitle{Li \MakeLowercase{\textit{et al.}}: \techNameNoSpace: A Knowledge Graph-Based Approach for Visualization Recommendation}

\abstract{
\revision{Visualization recommendation 
or automatic visualization generation 
can significantly lower the barriers for general users to rapidly create effective data visualizations,
especially for those users without a background in data visualizations. 
However, existing rule-based approaches require tedious manual specifications of visualization rules by visualization experts. Other machine learning-based approaches often work like black-box and are difficult to understand why a specific visualization is recommended, limiting the wider adoption of these approaches.}
This paper fills the gap by presenting \techNameNoSpace, a knowledge graph (KG)-based approach for visualization recommendation. It does not require manual specifications of visualization rules and can also guarantee good explainability.
Specifically, we propose a framework for building knowledge graphs,
consisting of three types of entities (i.e., data features, data columns and visualization design choices) and the relations between them, to model the mapping rules between data and effective visualizations.
\revision{A TransE-based embedding technique
is employed to learn the embeddings of both entities and relations of the knowledge graph from existing dataset-visualization pairs. Such embeddings intrinsically model the desirable visualization rules.}
Then, given a new dataset, effective visualizations can be inferred from the knowledge graph with semantically meaningful rules.
\revision{We conducted extensive evaluations to assess the proposed approach, including quantitative comparisons, case studies and expert interviews.} The results demonstrate the effectiveness of our approach.
} 


\newcommand{\revision}[1]{\textcolor{black}{#1}}
\newcommand{\revisioncr}[1]{\textcolor{black}{#1}}
\newcommand{\techName}{\textit{KG4Vis }}
\newcommand{\techNameNoSpace}{\textit{KG4Vis}}

\CCScatlist{ 
 \CCScat{K.6.1}{Management of Computing and Information Systems}%
{Project and People Management}{Life Cycle};
 \CCScat{K.7.m}{The Computing Profession}{Miscellaneous}{Ethics}
}

\teaser{
  \centering
  \vspace{1em}
  \includegraphics[width=\linewidth]{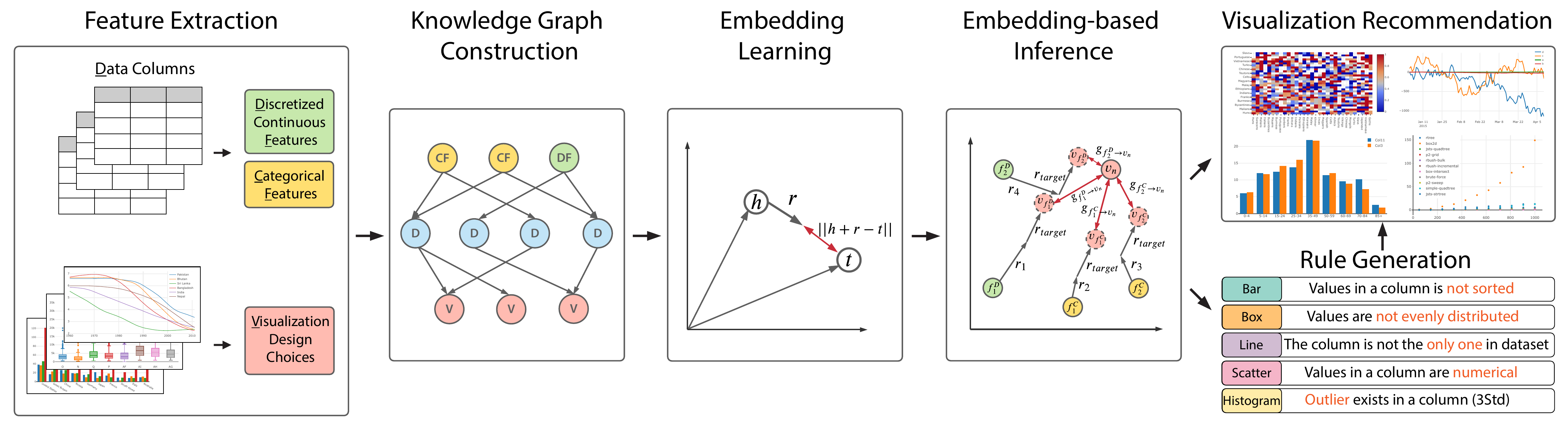}
  \caption{This figure illustrates the overall workflow of \techNameNoSpace. We extract features from existing dataset-visualization pairs and construct a knowledge graph~(KG). Then the embeddings of entities and relations in the KG are learned. Based on the embeddings, we conduct inference on a new dataset and finally recommend multiple visualizations. Also, various rules are extracted based on the embeddings and presented together with recommended visualizations to improve the interpretability of visualization recommendation.}
  \label{fig:teaser}
}

\vgtcinsertpkg

\begin{document}



\input{source/introduction}

\input{source/related_work}
\input{source/background}

\input{source/method}

\input{source/evaluation}
\input{source/discussion}
\nopagebreak
\acknowledgments{
This research was supported by the Singapore Ministry of Education (MOE) Academic Research Fund (AcRF) Tier 1 grant (Grant number: 20-C220-SMU-011).
We would like to thank the experts in our expert interviews and anonymous reviewers for their feedback.}

\bibliographystyle{abbrv-doi}

\balance

\bibliography{template}

\end{document}

%% file: source/introduction.tex
\firstsection{Introduction}

\maketitle

Data visualization
can effectively facilitate data exploration, insight communication and decision making
in various application domains such as business, scientific research, social media and journalism
~\cite{munzner2014visualization,ward2010interactive}.
However, transforming the input data to an effective visualization requires users to be familiar with both the input data and visualization tools.
Many existing visualization tools have a steep learning curve~\cite{data2vis2019dibia}. For example, the popular data visualization packages including D3~\cite{bostock2011d3}, Vega~\cite{satyanarayan2015reactive} and ggplot2~\cite{wickham2010layered} require users to know programming languages (e.g., JavaScript and R) as well as the syntax of these packages. 
Those methods are often tedious and time-consuming for generating visualizations.
Though dedicated visualization tools like Microsoft Excel or Google Spreadsheets are able to create standard charts by using templates, users still need to manually specify data attributes, and the mapping between them and the visual encodings. Also, these dedicated visualization tools offer limited expressiveness 
and customization.
To address these challenges, researchers have proposed a series of techniques and tools to automatically generate or recommend effective visualizations for input datasets~\cite{zhu2020survey,wang2020applying}.

Existing techniques and tools for automated visualization design and recommendation mainly consist of two categories~\cite{vizml2019hu}: rule-based approaches and machine learning (ML)-based approaches.
Rule-based approaches, such as APT~\cite{apt1986machinlay}, SAGE~\cite{roth1994interactive}, Show Me~\cite{showme2007mackinlay}, SeeDB~\cite{vartak2015seedb} and Foresight~\cite{demiralp2017foresight}, take into account the underlying data characteristics (e.g., statistical measures) and further leverage visualization principles or perceptual heuristics to automatically generate visualizations for analysts. 
Such rule-based approaches are also often augmented by supporting data variable selection and faceted browsing of recommended charts~\cite{voyager22017Wongsuphasawat,wongsuphasawat2015voyager}.
These rule-based approaches are straightforward and intuitive to understand.
However, they intrinsically suffer from limitations.
For example, these approaches need an explicit list of rules and heuristics that often rely on expert judgment. It is difficult and tedious to compile a complete rule list due to the necessity of significant manual effort.
Also, the rules may not be generalizable to different datasets or visualization choices. With the increase of the input data dimensions, there will be
explosive combinations
of visualization recommendations.

ML-based approaches (e.g., VizML~\cite{vizml2019hu}, Data2Vis~\cite{data2vis2019dibia}, 
and DeepDrawing~\cite{wang2019deepdrawing}) often train a machine learning model (especially a deep learning model) to directly learn the visualization rules from  examples of dataset-visualization pairs.
For instance, 
Data2Vis~\cite{data2vis2019dibia} leverages an LSTM-based neural translation model, 
which is trained on a Vega-Lite visualization corpus, to achieve the automatic generation of data visualizations.
There has been an increasing trend to apply deep learning techniques to visualization generation and recommendation~\cite{zhu2020survey,wang2020applying},
since they do not require users to manually specify the rules of visualization recommendation.
The contexts of the data and visualization are implicitly encoded by the trained models.
However, these deep learning models often work as a black box~\cite{hohman2018visual}. Thus, it is often difficult for users, 
to understand why a specific visualization is recommended, which further affects their trust in the recommended visualizations.

Motivated by the limitations of prior studies, we aim to achieve visualization recommendation that
requires \textbf{no manual specifications of visualization rules} and also guarantees \textbf{good explainability} of the recommendations to general visualization users.
It is a non-trivial task and little research has been conducted in this direction.

In this paper,
we propose \techNameNoSpace, a \underline{K}nowledge \underline{G}raph-based approach \underline{for} \underline{Vis}ualization recommendation. 
Similar to ML-based approaches for visualization recommendation, our approach also distills the rules for visualization design from existing examples such as the dataset-visualization pairs collected in VizML~\cite{vizml2019hu}.
Knowledge graphs are widely used in the natural language processing field for different purposes such as question answering and product recommendation~\cite{kagnet2019lin, cke2016zhang} and have shown great potential in modeling the relationship between different entities (i.e., knowledge). Such knowledge is often semantically understandable to human users.
Inspired by this, we explored how a knowledge graph can be constructed 
to model visualization rules
and further apply it to achieve effective visualization recommendations.
Specifically, we extract an extensive list of data features for each dataset and further build a knowledge graph based on three types of entities: \textit{data features}, \textit{data columns} and \textit{visualization design choices}. 
The directed links between two types of entities indicate different semantic meanings, e.g., \textit{``(One feature) is a data feature of (a specific data column)''}, \textit{``(One data column) is visually encoded by (a specific visualization design choice)''}.
Then, we employ TransE~\cite{transe2013bordes} to learn the embeddings of both relations and entities by using the dataset-visualization pair examples. 
These embeddings essentially encode the visualization generation rules
presented in
the examples.
Given a new dataset input, our approach can directly infer the relevant rules and recommend effective visualizations for the dataset, which intrinsically provides two desirable advantages.
First, users can trace back to the relevant rules to understand why a specific visualization is recommended, thereby enhancing their trust in the visualization recommendations (\textbf{explainability}) and improving their knowledge of visualization principles. 
Also, the whole recommendation process is data-driven and fully automated (\textbf{no manual specification of visualization rules}).

We investigated the effectiveness and usability of our approach through both quantitative and qualitative evaluations.
Specifically, to verify the knowledge graph model choice, we compared the visualization recommendation accuracy of knowledge graph models with that of other knowledge graph models.
Further, we conducted in-depth interviews with 12 visualization experts to assess whether our recommendation rules and recommended visualizations are meaningful to human users. We also showcased the visualization recommendation rules and the recommended visualizations to provide support for the effectiveness of {\techNameNoSpace}.

In summary, the major contributions of this paper are as follows:

\begin{compactitem}
\item We present \techNameNoSpace, a novel knowledge graph-based approach for visualization recommendation, which is essentially a data-driven approach and explainable to human users. To the best of our knowledge, this is the first time that a knowledge graph is employed to model visualization principles and recommend effective visualizations.

\item We conduct extensive evaluations, including qualitative comparisons with other models, case studies on different types of charts and in-depth expert interviews, which demonstrates the effectiveness and explainability of our approach.

\item We summarize the detailed lessons we have learned during the development of
{\techNameNoSpace},
which, we hope, can benefit subsequent work on applying knowledge graphs in the visualization field.

\end{compactitem}

%% file: source/related_work.tex
\section{Related Work}
Our related work can be categorized into three groups: visualization recommendation, knowledge graph-based recommendation and knowledge graph embedding.

\subsection{Visualization Recommendation}
\label{sec_visrec}
Visualization has been widely used for exploratory data analysis and decision making in various domains including stock trading~\cite{tradao2020tsang}, online education~\cite{proctoring2021li} and urban planning~\cite{topology2021feng}. However, most existing tools for creating visualizations heavily rely on users' manual specifications~\cite{vizml2019hu}.
To facilitate visualization usage by users with no background of visualization,
many researchers have recently explored the automated visualization recommendation using 
rule-based methods and machine learning (ML)-based methods~\cite{vizml2019hu
, wu2021mobile}.

Rule-based visualization recommendation methods are mainly based on manually-specified rules for mapping data to visual encodings
according to previous studies on human perception on visualizations~\cite{semiology2010bertin, graphical1984cleveland,assessing2018kim}. 
Representative studies include APT~\cite{apt1986machinlay}, SAGE~\cite{roth1994interactive}, Show Me~\cite{showme2007mackinlay} and Voyager2~\cite{voyager22017Wongsuphasawat}.
Though rule-based visualization recommendation methods have been widely studied 
their disadvantages are also obvious.
Since the rules are proposed or summarized by humans, the effort of constructing a thorough list of rules is enormous and the following update of rules can be hard~\cite{wide&deep2020qian}. 

On the contrary, ML-based approaches have been investigated for visualization recommendation to address the limitation of rule-based methods. 
For example,
DeepEye~\cite{deepeye2019luo} and Draco~\cite{draco2019moritz} have been proposed to augment existing rules proposed by experts with learning the preference of visualizations and then ranking recommended visualizations. 
Hu~\textit{et al.}~\cite{vizml2019hu} and Qian~\textit{et al.}~\cite{wide&deep2020qian} fed features of datasets to Neural Networks~(NNs) to infer how the datasets are represented by visualizations.
These models have advanced the performance of visualization recommendations.
However, these methods often use a deep learning approach and the ``black-box'' nature of these models makes it hard to
interpret the recommended results. 

In this paper, we aim to propose a visualization recommendation method based on a knowledge graph to leverage the advantages of both rule-based and ML-based methods. It can recommend satisfactory visualizations and also make the reasons behind the recommendations transparent to users by providing explainable recommendation rules.

\subsection{Knowledge Graph-based Recommendation}\label{sec:kg-rec}
Incorporating \textit{knowledge graphs}~(KGs) in recommender systems is an emerging research direction in recent years~\cite{surveyKGRec2020guo}.
KGs are structured representations of human knowledge and consist of entities and relations between entities~\cite{surveyKG2020ji}.
KG-based recommendation algorithms have two major advantages:
effectively modeling different latent relations between entities
~\cite{surveyKGRec2020guo} and providing explainable recommendation results based on the graph structure~\cite{ripplenet2018wang}. 
According to Guo~\textit{et al.}~\cite{surveyKGRec2020guo}, there are mainly three types of methods for KG-based recommender systems, embedding-based methods~\cite{bem2019ye, cke2016zhang}, path-based methods~\cite{metagraph2017zhao, pgpr2019xian} and unified methods~\cite{ripplenet2018wang, kgat2019wang}.

Among these three types of methods, our recommendation algorithm is closer to embedding-based methods. 
We explicitly learn the embeddings of entities and relations in our KG and conduct recommendations based on the embeddings.

\subsection{Knowledge Graph Embedding}\label{sec:KGE}
Knowledge graph embedding~(KGE) represents the entities and relations in a KG with low-dimensional embedding vectors so that various entities and relations can be easier to handle in downstream tasks~\cite{wang2017knowledge}, such as link prediction~\cite{rotate2019sun, transh2014wang} and triplet classification~\cite{transe2013bordes, transh2014wang}.

According to prior studies~\cite{wang2017knowledge,hogan2020knowledge},
KGE methods can mainly be categorized into 3 classes: translational models, tensor decomposition models and neural models.
Translational models aim to model the relations between two entities as a translation in space.
The most representative method under this category is TransE~\cite{transe2013bordes}.
It assumes that the embedding of a tail entity should be the sum of the embedding of a head entity and a translation vector which is the relation between them.
A more detailed introduction to TransE is provided in Section~\ref{sec:background}.
Following TransE, a series of translational methods have been proposed.
For example, TransR~\cite{transr2015lin} 
represents relations in different spaces and conducts translation with a relation after projecting entities to the corresponding space.
RotatE~\cite{rotate2019sun} employs rotation in the complex space to represent relations between entities.
Another class of KGE models aims to extract the embeddings of entities by applying tensor decomposition to model the graph structures~\cite{hogan2020knowledge}, for example, RESCAL~\cite{rescal2011nickel} and DistMult~\cite{distmult2015yang}.
Recently, many neural models have also been proposed for KGE including SME~\cite{sme2014bordes} and ConvE~\cite{conve2018dettmers}.

In our paper, 
due to its efficiency and intuitiveness,
an improved version of the widely recognized embedding approach (i.e., TransE~\cite{transe2013bordes}) is applied in our model to learning the embedding vectors of entities and relations in our KG. 

%% file: source/background.tex
\section{Background: TransE}\label{sec:background}
As introduced in Section~\ref{sec:kg-rec}, a KG represents human knowledge as a directed graph consisting of entities and relations. 
Each entity is represented as a node in the graph and each relation is \revision{an} edge type.
In the directed graph, each edge indicates 
the existence of a relation between
two entities, i.e., the head entity and the tail entity.
Thus, it is common to use a triplet, $(head\: entity, relation, tail\: entity)$, to represent an edge, which can also be denoted as $(h, r, t)$ in short.
However, if a KG is represented with symbolic triplets, the manipulation can be hard~\cite{wang2017knowledge}.
Thus, researchers propose knowledge graph embedding~(KGE) to represent entities and relations in KGs as continuous embedding vectors. 
TransE~\cite{transe2013bordes} is one of the most representative KGE methods with great efficiency and intuitiveness. 
In this paper, we leverage an approach based on TransE to learn the representations of various data features and visualization designs, as well as the relations between data features and visualization designs, which are further used to recommend appropriate visualizations.
The basic idea of TransE is that the relation $r$ between a head entity $h$ and a tail entity $t$ can be approximately represented by a translation from $h$ to $t$. Suppose the embedding vectors of $h, r, t$ are $\mathbf{h},\mathbf{r}, \mathbf{t}$, the relationship among them can be written as 
\begin{equation}\label{equ:embedding_relationship}
    \mathbf{h} + \mathbf{r} \approx \mathbf{t},
\end{equation}
which is also illustrated in Fig.~\ref{fig:inference}(a).
Thus, the scoring function that measures the possibility of a triplet's existence is defined as
\begin{equation}\label{equ:score}
    g(h, r, t) = -||\mathbf{h}+\mathbf{r}-\mathbf{t}||_{1/2},
\vspace{-0.5em}
\end{equation}
where a larger score indicates that the triplet $(h, r, t)$ tends to be plausible and the distance can be calculated by using either L1 or L2 distance.
With the scoring function of triplets, TransE applies a margin-based ranking criterion as the loss function in embedding learning:
\begin{equation}
    L = \sum_{{(h' ,r, t') \in S'}}ReLU(\gamma+g(h', r, t')-g(h, r, t)),
\vspace{-1em}
\end{equation}
where $\gamma > 0$ is a margin parameter and $S$ and $S'$ denotes the set of training triplets and the set of negative triplets, respectively.
Intuitively, we can consider the loss function which aims to make the difference between scores of training triplets and scores of negative triplets as large as possible.
Here the negative triplets are generated by a random replacement of the head entity or the tail entity in a training triplet.
In the training process, TransE applies gradient descent to minimize the loss and optimize the learned embeddings.

%% file: source/method.tex
\section{Method}
\techName 
consists of four major modules: feature extraction, KG construction, embedding learning, and embedding-based inference, as illustrated in Fig.~\ref{fig:teaser}.
First, we extract both data features and visualization design choices from dataset-visualization pairs~(Section~\ref{sec:feature}).
Then, we propose building knowledge graphs to model the mapping from datasets to visualizations.
Specifically, we define three types of entities (i.e., data features, data columns and visualization design choices) and the relations between them~(Section~\ref{sec:kg_construction}).
Further,
we leverage the classic knowledge graph embedding approach \textit{TransE} to represent the entities and relations with embedding vectors~(Section~\ref{sec:embedding}).

The final step of \techName is to infer visualization design choices for a new dataset, where explicit visualization rules are also generated~(Section~\ref{sec:inference}).

\subsection{Data Features and Visualization Design Choices}\label{sec:feature}

Inspired by VizML~\cite{vizml2019hu}, we also extract \textit{data features} of a dataset by quantifying the characteristics of its individual columns, where each column of the dataset is referred to as a \textit{data column} in this paper.
Specifically, we extract 81 data features introduced in VizML~\cite{vizml2019hu}.
These features include 50 continuous features and 31 categorical features.
The data features of data columns are categorized into three classes:
the data type of the data column~(Types), the statistical features of values in the data column such as distribution and outliers~(Values) and the name of the data column~(Names).
A detailed feature list is available in the supplementary material.

To visualize a given data column, 
the key \textit{visualization design choices} are the type of the visualization and the axes where each data column is encoded. 
In our method, we limit our scope of recommended visualizations to six common types of 2-D visualizations: \textit{bar charts, box plots, heatmaps, histograms, line charts} and \textit{scatter plots}.
In the rest of the paper, we will use \textit{bar, line, scatter, box, heatmap}, and \textit{histogram} to refer to them respectively for simplicity.
Since these visualization types supported in our method are 2-D visualizations, the axes where the data is encoded are limited to horizontal and vertical axes, which are denoted as \textit{x-} and \textit{y-axis} in the rest of our paper.

\subsection{Knowledge Graph Construction}
\label{sec:kg_construction}

\begin{figure}[h!]
    \centering
    \includegraphics[width=\linewidth]{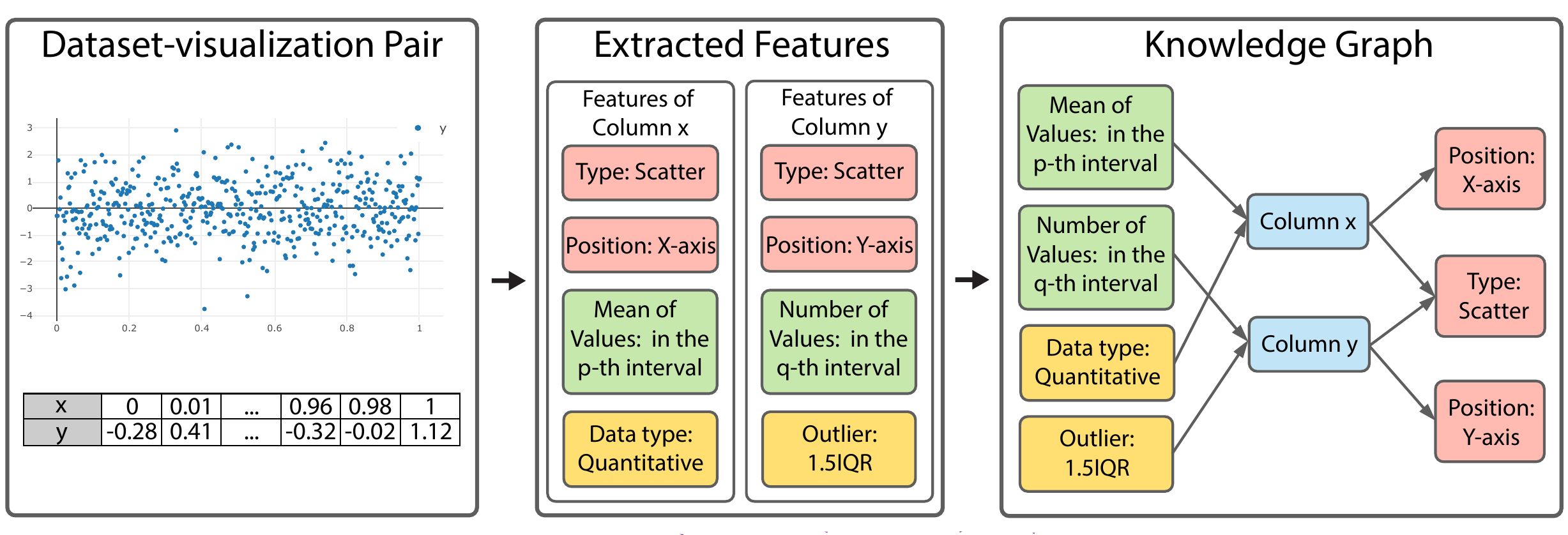}
    \caption{\revision{This figure shows an example of the transformation from a dataset-visualization pair to a part of the KG. Only a part of features and entities are shown. Red nodes represent visualization design choices. Blue nodes represent data columns. Green and yellow nodes represent discretized data features and continuous data features, respectively.
    }
}
    \vspace{-1em}
    \label{fig:kg_construction}
\end{figure}

We build knowledge graphs (KGs) to model the complex mapping from data to visualizations.
According to our survey, there are no existing studies on constructing knowledge graphs for visualization recommendations.
We propose defining four types of entities and three types of relations among entities to delineate the mapping from a dataset to appropriate visualizations.
By leveraging a large number of existing dataset-visualization pairs in real practice, our knowledge graphs can capture the widely-used visualization design principles.
An overall description of the knowledge graph construction is shown in Fig.~\ref{fig:teaser} and an example is shown in Fig.~\ref{fig:kg_construction}.

\textbf{Definition of Entities.} 
The first step of building our KG is to define the entities.
Since 
the purpose of KG
is to represent the relationship between data features, data columns and visualization design choices, they are naturally considered as entities in the KG.
Generally, an entity in KG represents a categorical value, for example, a person or a place.
Thus, a data column, a visualization design choice or a categorical data feature 
can be directly
represented as an entity.
However, defining the entities for continuous data features is challenging, since continuous features are hard to be handled by KG.
One possible method is to assign each concrete value of a continuous data feature to an entity, for example, we create two entities to represent that the length of a data column is 5 and the length of a data column is 6, respectively.
However, the number of entities will inflate and then the KG will become sparse since the number of possible concrete values is large and the number of edges between an entity of a value and entities of data columns will be small. 
Learning the embeddings of a sparse KG with an enormous number of entities can consume lots of computational power and the learned embeddings can be of low quality~\cite{sparse2020zhao}.
Thus, we propose to divide each continuous data feature into different intervals,
and regard each discretized continuous data feature with a specific value interval as an entity in the KG.

There are many commonly used discretization strategies such as KMeans discretization, uniform discretization and quantile discretization. 
However, they suffer from a common pitfall: 
these discretization methods only consider the distribution of values of all data points and cannot take extra information~(e.g., predictive attributes) into consideration.
Thus, inspired by Ming~\textit{et al.}~\cite{rulematrix2019ming}, we apply a discretization method based on the minimum description length principle~(MDLP)~\cite{mdlp1993fayyad} to discretize the continuous data features.
With MDLP, the visualization type is able to be used as extra criteria when discretizing continuous data features. 
Besides the consideration of visualization types, it can automatically decide how many intervals are generated based on the distribution of data and the minimum size of each interval.
This is another advantage of MDLP, especially compared with other approaches like KMeans, where the number of intervals should be explicitly given.
With the discretized continuous data features generated by MDLP, we treat each interval of a feature as an entity.
In summary, we overall have four classes of entities, visualization design choices~($\mathbb{E}_V$), discretized continuous data features~($\mathbb{E}_{DF}$), categorical data features~($\mathbb{E}_{CF}$) and data columns~($\mathbb{E}_D$).
The detailed list of entities is available in Table~\ref{table:relations}.

\textbf{Definition of Relations.} 
After introducing four classes of entities, 
we further defined three classes of relations as shown in Table~\ref{table:relations}.
First, we have a class of relations that connect data columns to visualization design choices~($\mathbb{R}_{D \rightarrow V}$) and represent \textit{``(one data column) is visually encoded with (a specific visualization design choice)''}.
To be specific, 
there are two relations that belong to this class:
1) a data column (often together with other data column(s)) is visualized as a specific visualization type, e.g., bar; 
2) a data column is encoded on the x-axis or y-axis.
The complete list of this type of relations and corresponding entities is in the first section of Table~\ref{table:relations}.
The second and third classes of relations both describe the mapping from data features to data columns.
Thus, the semantic meaning of relations in these two \revision{classes} can be described as \textit{``(one feature) is a data feature of (a specific data column)''}.
The second class of relations are those linking categorical data features to data columns~($\mathbb{R}_{CF \rightarrow D}$).
For this class of relations, we group the categorical data features according to their semantic meanings and construct 13 relations, for example, \textit{``the general data type of (one data column) is (categorical)''}.
A detailed list of relations in this class is in the second section of Table~\ref{table:relations}.
The third class of relations 
aim to model the mapping 
from discretized continuous data features to data columns~($\mathbb{R}_{DF \rightarrow D}$). 
We define a relation for each continuous feature and there are 50 relations in total, which is equal to the number of continuous features as described in Section~\ref{sec:feature}.
An example is \textit{``the number of unique values in (one data column) is (between 0 and 30)''}.

After defining the entities and relations, we 
extract triplets from existing dataset-visualization pairs.
These triplets serve as the edges that link different entities to form a graph.
For a data column $d$ with its data feature set $F_d$, a set of triplets are extracted as 
$\{(f_i, r_i, d)|f_i \in F_d\}$, where $r_i \in \mathbb{R}_{CF \rightarrow D} \cup \mathbb{R}_{DF \rightarrow D}$ is the corresponding relation 
from $f_i$ to $d$.
Similarly, for the set of visualization design choices $V_d$ of the data column $d$, we extract another set of triplets $\{(d, r_n, v_n)|v_n \in V_d\}$.
Thus,
all the triplets associated with the data column $d$ can be denoted as
$\{(f_i, r_i, d)|f_i \in F_d\} \cup \{(d, r_n, v_n)|v_n \in V_d\}$.
By combining the triplets associated with all the data columns, we gain the knowledge graph that will be used for visualization recommendation.

\begin{table}[hbt!]
    \centering
    \small
    \caption{This table shows the definition of relations and their related entities. The first section presents $\mathbb{R}_{D\rightarrow V}$ and \revision{$\mathbb{E}_{V}$}. The second section of the table illustrates $\mathbb{R}_{CF\rightarrow D}$ and \revision{$\mathbb{E}_{CF}$}. The third section shows $\mathbb{R}_{DF\rightarrow D}$ and \revision{$\mathbb{E}_{DF}$}.
    IQR is the interquartile range
    and Std is the standard deviation.} 
    \setlength{\aboverulesep}{0.5pt}
    \setlength{\belowrulesep}{0.5pt}
    \begin{tabular}{p{0.6cm}|p{3.4cm}|p{3.3cm}}
        \toprule
        Class & Relations & Corresponding Entities \\
        \midrule
        \multirow{2}{*}{$R_{D \rightarrow V}$}& Visualization type of the column is & bar, box, heatmap, histogram, line, scatter \\\cmidrule{2-3}
        &The column is encoded on & x-axis, y-axis \\\midrule
        \multirow{7}{*}{$R_{CF \rightarrow D}$}&
        The general data type \revision{of the data in the column} is & categorical, quantitative, temporal\\\cmidrule{2-3}
        &The specific data type \revision{of the data in the column} is & string, integer, decimal, datetime \\\cmidrule{2-3}
        &The name of the data column contains & ``x'', ``y'', ``time'', digit, whitespace, ``\$'',``€'', ``£'', ``\yen''\\\cmidrule{2-3}
        &Outlier exists in the column according to criteria & 1.5IQR rule, 3IQR rule, 3Std rule, (1\%, 99\%) rule \\\cmidrule{2-3}
        &Values are normal at & $(p<0.01)$, $(p<0.05)$ \\\cmidrule{2-3}
        &Values are sorted/are monotonic/ are in linear space/are in log space/ are unique, missing value is in the column, the column is the only column in dataset & \{\textit{relation}\} is true, \{\textit{relation}\} is false ~(\textit{relation} refers to a relation in the left.)\\\cmidrule{2-3}
        &The column name starts with & upper case, lower case \\
        \midrule
        \multirow{2}{*}{$R_{DF \rightarrow D}$}&
        Continuous features (e.g., number of unique values in the column) & Value of the continuous feature is in the 1st, ..., $m$-th interval~($m>1$)~(e.g., number of unique values in the column is in the 4th interval)\\

        \bottomrule 
        
    \end{tabular}
    \label{table:relations}
\end{table}

\subsection{Embedding Learning}\label{sec:embedding}
In this section, learning to represent entities and relations by embedding vectors with TransE is introduced.
The major advantage of using embeddings to represent entities and relations is that the manipulation of them in KG~\cite{wang2017knowledge} is convenient, which greatly benefits our subsequent inference and explicit rule generation. 

In our approach, 
we adopt an improved TransE with self-adversarial negative sampling~(denoted as TransE-adv)~\cite{rotate2019sun} due to its efficiency and intuitiveness~\cite{wang2017knowledge}.
The major advantage of adopting self-adversarial negative sampling is that it can effectively improve the learning efficiency by considering the current embedding model and eliminating obviously false triplets.
According to Sun~\textit{et al.}~\cite{rotate2019sun}, each negative sample is assigned a weight which indicates its probability of being true when using the current embeddings. 
The weight is calculated as 
\begin{equation}\label{equ:negative_sampling}
w\left(h^{\prime}, r, t^{\prime}\right)=\frac{\exp (\alpha g(h', r, t')) }{\sum_{(h_i' ,r, t_i') \in S'} \exp (\alpha g(h_i', r, t_i'))},
\end{equation}
where $(h', r, t')$ is a negative sample, $S'$ is the collection of negative samples, $\alpha$ is the temperature of sampling~\cite{temperature1985ackley} and $g$ is the scoring function.
Based on the weight of negative samples, a negative sampling loss function is applied in the training of TransE-adv, which is as follows:
\begin{equation}\label{equ:negative_sampling_loss}
\begin{split}
L= & -\log \sigma\left(\gamma+g(h, r, t)\right)\\
 & -\sum_{{(h' ,r, t') \in S'}} w\left(h^{\prime}, r, t^{\prime}\right) \log \sigma\left(-g(h', r, t')-\gamma\right),
\end{split}
\vspace{-1em}
\end{equation}
where $\sigma$ is the Sigmoid function and $\gamma$ is the margin.

The overall procedure of embedding learning in our approach is as below. First, the embeddings of entities and relations are initialized.
Then in each step, we randomly sample a batch of training triplets and generate negative samples accordingly.
Then a weight value is assigned to each negative sample according to Equation~\ref{equ:negative_sampling}.
With the batch of training triplets and negative triplets, the loss of current embeddings is calculated as Equation~\ref{equ:negative_sampling_loss}.
Finally, the loss is used for the optimization of embeddings in each step.

\input{source/inference}

%% file: source/inference.tex
\subsection{Inference with Embeddings}\label{sec:inference}
In the previous step, the embeddings of entities and relations in the KG are learned.
In this section, we introduce the method of leveraging these embeddings to infer how to visually encode a new data column which has multiple data features.

As indicated by Equation~\ref{equ:embedding_relationship}, the tail entity's approximate embedding equals the sum of the embedding of a given head entity and the embedding of the relation between them.
Based on this assumption, we propose a method to infer the final visualization design choice of a data column in a new dataset.
We first
extract rules indicating the mapping from data features to visualization design choices.
Then, we aggregate all the rules to decide the final visualization design choices.
Here each rule has a structure like ``\textbf{If} \textit{the data column has \textbf{\{a data feature\}}}, \textbf{then} \textit{the data column can be represented by \textbf{\{a visual design choice\}}}'', where the data feature refers to a categorical feature or a discretized continuous feature.
For the rest of this paper, each rule will be denoted as ``\textit{\textbf{a data feature}} $\rightarrow$ \textit{\textbf{a visual design choice}}'' for simplicity.
In each rule, the part before ``$\rightarrow$'', i.e.,  ``\textit{\textbf{a data feature}}'', is also called the condition of the rule.
To derive such kind of rules from KG with embeddings, for each data feature entity $f_{i} \in \mathbb{E}_{DF} \cup \mathbb{E}_{CF}$, we conduct translations twice. 
The first translation is to approximate the embedding of an imaginary data column $d_{im}$ which is only connected by $f_i$. 
The embedding of $d_{im}$ is computed as 
$\mathbf{f}_{i}+\mathbf{r}_{j}$, where $\mathbf{r}_{j}$ is the embedding of the relation connects $f_{i}$ to $d_{im}$.
The second translation is to infer how $d_{im}$ will be visually represented under a specific relation $r_{target} \in \mathbb{R}_{D \rightarrow V}$ by computing $\mathbf{f}_{i}+\mathbf{r}_{j}+\mathbf{r}_{target}$.
After two translations, we can extract a rule as $f_{i} \rightarrow v_{n}$ with a score  defined as follows:
\begin{equation}\label{equ:score_rule}
    g_{f_{i} \rightarrow v_{n}}=-||\mathbf{f}_{i}+\mathbf{r}_{j}+\mathbf{r}_{target}-\mathbf{v}_{n}||,
\end{equation}
which indicates how much $\mathbf{v}_{n}$ is preferred given $\mathbf{f}_{i}$.
More specifically, the rule also reflects the possibility that $f_{i} \rightarrow v_{n}$ is correct. 
Then, by using the rules, we can calculate the average score of 
all the relations that link
a data feature $f_i \in F_{new}$ to a design choice $v_n$ as follows:

\begin{equation}\label{equ:score_featovis}
\begin{split}
    g(d_{new}, r_{target}, v_n) & = \frac{1}{|F_{new}|}\sum_{f_i\in F_{new}}g_{f_{i} \rightarrow v_{n}}. 
\end{split}
\end{equation}
\revision{This computation can also be viewed as an aggregation of all rules whose conditions are fulfilled by the current data column.
Then, for every visualization design choice, the aggregation is computed to get the score of recommending it.
}
After obtaining scores of all possible design choices, we compare them and select the visualization choice with the highest score as the inferred visualization choice for $d_{new}$.
For example, when we decide whether $d_{new}$ should be encoded on which axis,
$g(d_{new}, r_{axis}, v_{x-axis})$ and $g(d_{new}, r_{axis}, v_{y-axis})$ can be computed with Equation~\ref{equ:score_featovis}. If $g(d_{new}, r_{axis}, v_{y-axis}) > g(d_{new}, r_{axis}, v_{x-axis})$, then we recommend to encode $d_{new}$ on y-axis.

After inferring the visualization type and the axis that will be used for each data column, we need to
collectively consider the inference results of all the columns of a dataset and further assemble valid visualizations for the dataset.
Thus, we need to propose a set of post-processing rules according to the visualization grammar~(e.g., Vega-lite and Plotly)
to guarantee that valid visualizations are generated.
For example, in Plotly, among all the six visualization types, box plots and histograms only require specifications on one axis while other types requires specifications on both x-axis and y-axis.
Thus, if any data column in a dataset is inferred to be visualized in histogram or box plots, all columns will be visualized on either x- or y-axis.
Otherwise, we plot all columns according to their inferred axes.
Since our method will assign a score to different visualization choices~(e.g., bar and line), it is possible for us to recommend several different visualizations with top $k$ visualization choices according to their scores.

\begin{figure}[ht!]
    \centering
    \includegraphics[width=0.9\linewidth]{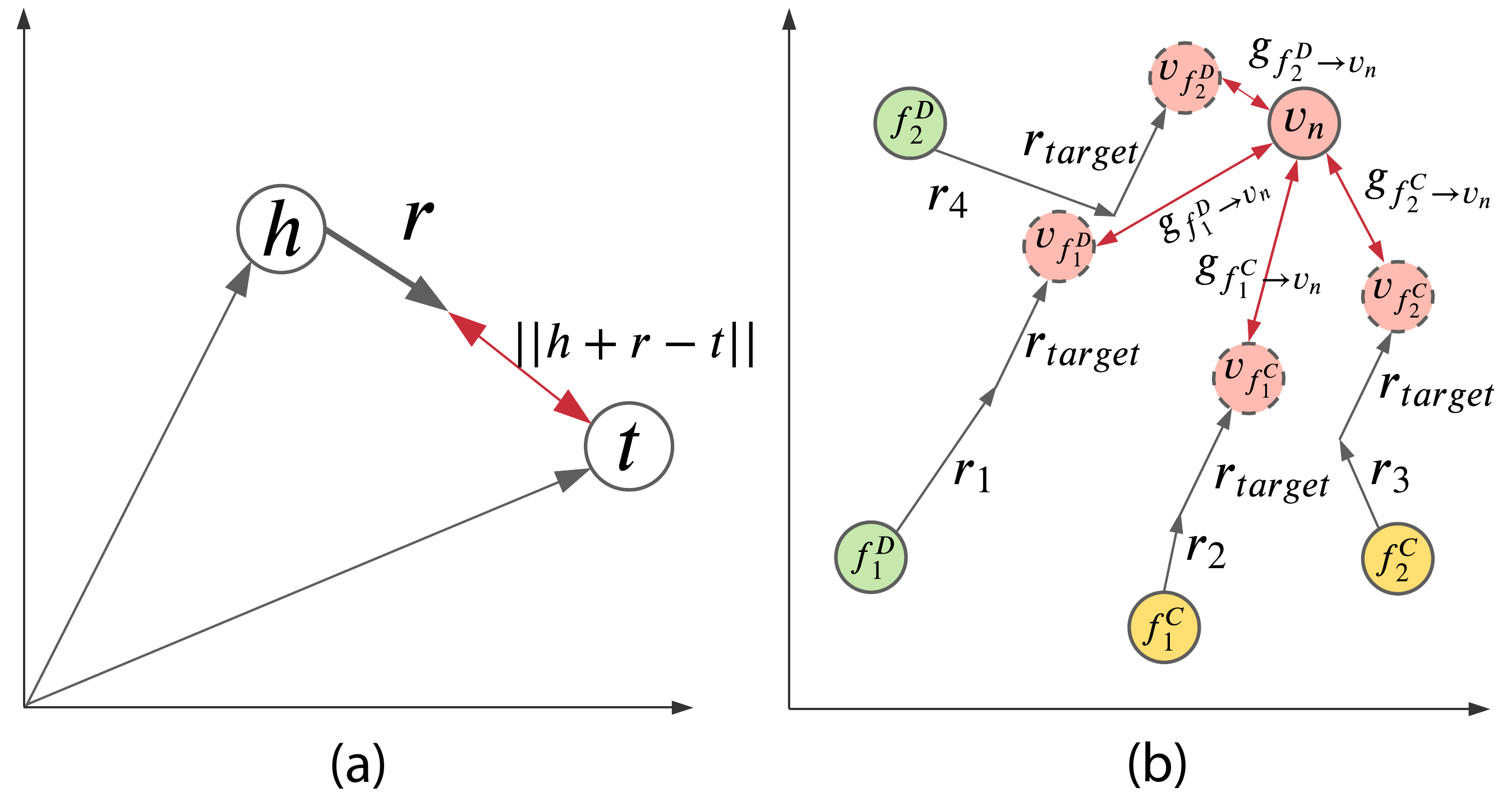}
    \vspace{-1em}
    \caption{This figure illustrates (a) TransE in 2-D space and (b) our inference method. In (a), two points, $h$ and $t$, represent the embeddings of head and tail entities while vector $r$ denotes the embedding of relation between two entities. The red arrow with two heads shows the distance $||\mathbf{h}+\mathbf{r}-\mathbf{t}||$ between $\mathbf{h}+\mathbf{r}$ and $\mathbf{t}$.
    In (b), each point is a 2-D embedding of an entity.
    $f^D_1$ - $f^D_2$ and $f^C_1$ - $f^C_2$ represent two discretized continuous data features and two categorical data features of a new data column.
    $r_1$ - $r_4$ are relations connecting data features to data columns.
    $v_n$ is a visualization design choice connected by $r_{target}$.
    $v_{f}$s are the estimated embeddings of visualization design choices.
    Red arrows with two heads denote the distances between embeddings. $g$ represents the scoring function.}
    \label{fig:inference}
\end{figure}

%% file: source/evaluation.tex
\section{Evaluation}
\label{sec-evaluation}

This section introduces the evaluation of our method from the following perspective: evaluation setup, quantitative evaluations and qualitative evaluations.
Qualitative evaluations consist of case studies, expert interviews and a comparison between generated and empirical rules. 
The source code of \revision{{\techNameNoSpace}}, the visualization corpus used in the evaluations and the supplementary material are available in \textcolor{navy}{\url{https://kg4vis.github.io/}}.

\subsection{Evaluation Setup} \label{sec:evaluation_setup}
The evaluation settings include the generation of our visualization corpus, the pre-processing and hyper-parameter setting for embedding learning, and the visualization generation settings.

\textbf{Visualization Corpus.} We used
the VizML corpus~\cite{vizml2019hu} for our evaluations. It contains around 120k dataset-visualization pairs. 
In this corpus, each visualization is generated by Plotly in JavaScript.
According to VizML~\cite{vizml2019hu}, users may slightly modify their datasets and create duplicated visualizations, so they randomly sample one dataset of each user to generate a corpus for evaluation.
We followed the same sampling strategy of VizML and created a new corpus for our evaluation, which covers the
datasets of six visualization types (i.e., bar, box, heatmap, histogram, line and scatter).
Note \textit{dataset} in this paper refers to a table consisting of multiple columns. 
The final corpus has 88,548 dataset-visualization pairs and 309,335 data columns in total.

\begin{figure*}[!hbt]
    \centering
    \includegraphics[width=0.8\linewidth]{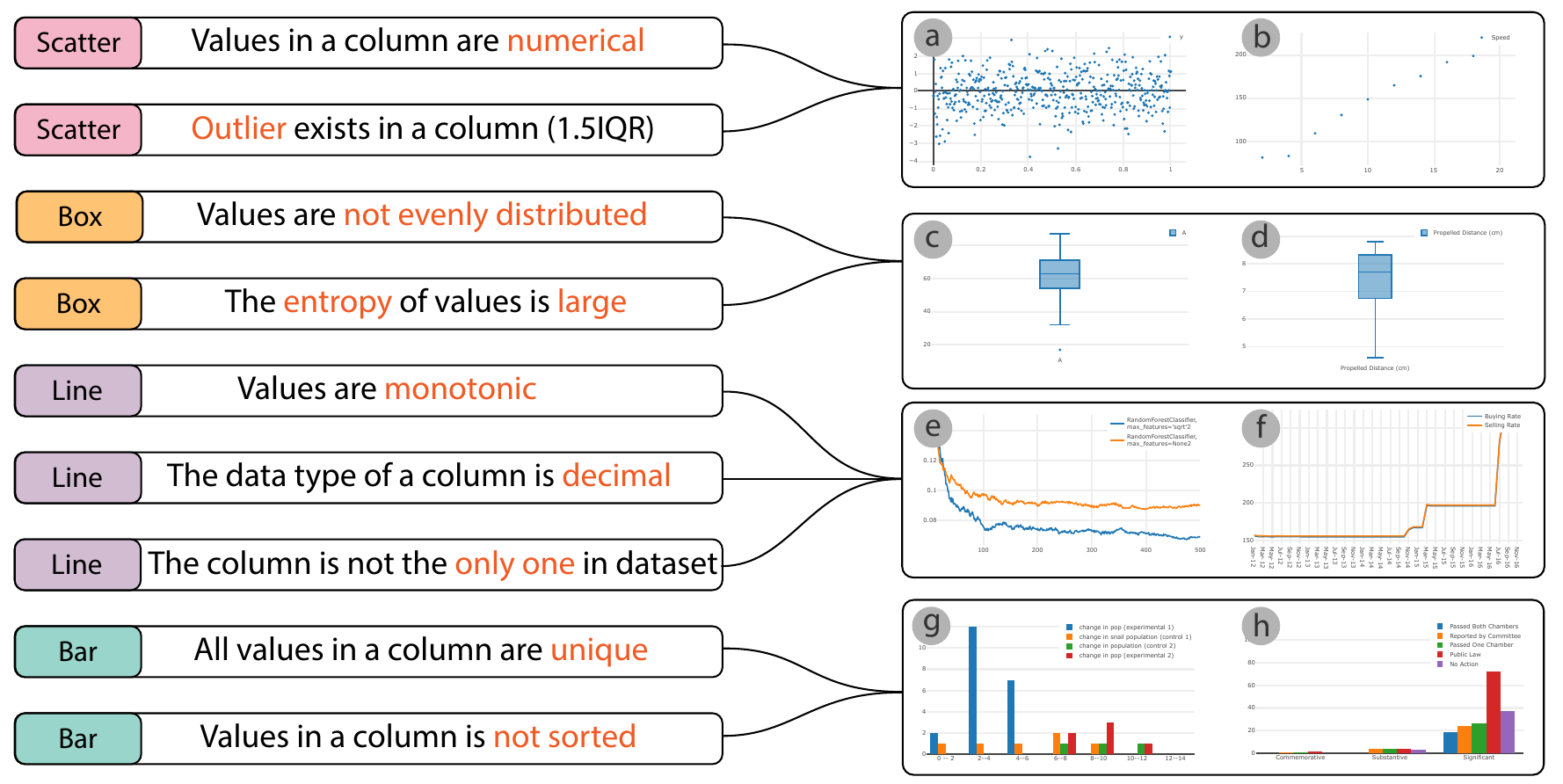}
    \caption{This figure shows a gallery of recommended visualizations in (a)-(h) and related rules with high scores. The listed rules are only a small subset of all rules applied in the recommendation of corresponding visualizations. Due to limited space, we only present rules of 4 visualization types, bar charts, box plots, line charts and scatter plots.}
    \vspace{-2em}
    \label{fig:gallery}
\end{figure*}

\textbf{Embedding Learning Setup.}
To learn the embeddings of entities and relations, we randomly selected 70\% of our corpus as the training set.
Then, we extracted features and conducted several steps of pre-processing. 
First, we removed invalid data columns, e.g., data columns without a visualization type.
Then, we dealt with extreme values by
using the 5\% quantile and the 95\% quantile of each continuous feature to replace values
less than the 5\% quantile or larger than the 95\% quantile respectively.
After pre-processing, a KG can be constructed on the training set by extracting entities, relations and triplets, as introduced in Section~\ref{sec:kg_construction}\revision{.}
To extract entities representing continuous data features, MDLP was applied to discretize each continuous data feature to several intervals.
When applying MDLP, we set the minimum proportion of samples to split an interval to $0.1$ and the minimum proportion of samples in an interval to $0.05$~(i.e., at most each continuous feature can be split into 20 intervals).
The reason why we selected these two values is that we need to strike a balance between performance and intuitiveness.
When there are more intervals for a data feature, the recommendation results may be better.
However, an excessive number of intervals for a data feature can 
be harmful to the interpretability of rules.
Thus, we adopted the values above
according to our empirical observation.

In our KG, there are in total 216,851 entities, 56 distinct relations and 9,679,463 triplets.
Then, TransE-adv was applied on the KG to learn the embeddings for inference.
All the embedding vectors of entities and relations have 1,000 dimensions following a previous study~\cite{rotate2019sun} and were initialized using a uniform distribution.
In the training process, the batch size of each epoch was 1,024
and the number of training steps was 30,000.
The initial learning rate was set as $0.001$.
Following the study by Sun~\textit{et al.}~\cite{rotate2019sun}, we chose Adam~\cite{adam2015kingma} as the optimizer
and L2 distance was applied in all the scoring functions in Equations~\ref{equ:negative_sampling} -  \ref{equ:score_featovis}.
\nopagebreak

\subsection{Quantitative Evaluation}\label{sec:quantitative_results}
To evaluate our embedding learning method, we conducted experiments. 
As stated in Section~\ref{sec:feature}, we have two major tasks: 1) the inference of visualization types and 2) the inference of axis to be encoded on for a data column.
To make our evaluation reliable, we conducted a 5-fold cross-validation and report results of different methods.

\textbf{Metrics.} 
In our inference method, each visualization design choice is given a score and then ranked in descending order.
Thus, to comprehensively evaluate our method, we utilized two widely used metrics, the average rank of correct design choices~(denoted as \textit{MR}) and the proportion of correct visualization design choices ranked in the top two inferred design choices~(denoted as \textit{Hits@2})~\cite{transe2013bordes} to evaluate the performance of visualization types.
Since the inference on the axis is binary~(i.e., either x- or y-axis), we evaluated it by accuracy.

\textbf{Baseline Models.}
The baseline embedding learning models used in our quantitative evaluation were TransE and RotatE~\cite{rotate2019sun}.
TransE here refers to TransE without self-adversarial negative sampling.
The reason why we selected TransE without self-adversarial negative sampling as a baseline is that we would like to confirm self-adversarial negative sampling can advance the performance of embedding learning by comparing TransE-adv with TransE\revision{.}
Another baseline model is RotatE.
By comparing TransE-adv with RotatE, we would like to make sure TransE-adv can achieve a satisfactory performance compared to one of the state-of-the-art KGE models.
As introduced in Section~\ref{sec:KGE}, RotatE models the relationship between two entities as a rotation in the complex space and has a different scoring function which is defined as $g(h, r, t)= -\|\mathbf{h} \circ \mathbf{r}-\mathbf{t}\|$,
where $\circ$ indicates element-wise product.
The inference method of \techName can still apply to the embedding learned by RotatE 
by replacing the scoring function of TransE with that of RotatE.

\textbf{Results.}
According to the results in Table~\ref{table:quantitative_evaluation}, TransE-adv outperforms other embedding learning models on both visualization type inference and axis inference.

\begin{table}[h!]
\centering
\caption{This table shows the result of our quantitative evaluation on embedding learning methods. The best results are in bold. Overall, TransE-adv outperforms others. Among all metrics, a smaller MR indicates better performance while larger accuracy and Hits@2 are better.}

\small
\centering
\setlength{\aboverulesep}{0.5pt}
\setlength{\belowrulesep}{0.5pt}
\begin{tabular}{llll}
\toprule
 &   Axis & \multicolumn{2}{l}{Visualization Type} \\ \cmidrule{2-4} 
 &   Accuracy & MR & Hits@2 \\ \midrule \midrule
\multirow{1}{*}{TransE-adv} 
& \textbf{0.7350} & \textbf{1.9567} &  \textbf{0.7489} \\ \midrule
 \multirow{1}{*}{TransE} & 0.7214  & 1.9718 &  0.7445\\ \midrule
\multirow{1}{*}{RotatE} & 0.7193  & 1.9608 & 0.7458 \\
\bottomrule
\end{tabular}
\label{table:quantitative_evaluation}
\vspace{-1em}
\end{table}

\subsection{Qualitative Evaluation}

In this section, we evaluate \techName by conducting case studies, expert interviews and comparing the generated rules with empirical rules.
\subsubsection{Case Study}\label{sec:case_study}

Fig.~\ref{fig:gallery} shows several recommendation results and rules to illustrate the effectiveness of our approach and how rules can guide visualization recommendation.
These recommended visualizations and rules were rated highly in our expert interviews.
Rules in Fig.~\ref{fig:gallery} are related to four types of visualization: scatter, box, line and bar.
\revisioncr{To make the original rules more understandable for general users, we translated those rules with complex statistical meanings 
to straightforward descriptions. 
For example, \textit{``the entropy of values is large $\rightarrow$ box''} is used to describe that box plots are often used when the entropy of the data column values falls in the discretized interval  
that corresponds to a large entropy.
When dealing with the rules with simple categorical features, we followed the 
structure introduced in Section~\ref{sec:inference} to generate rules such as \textit{``values in a column are numerical $\rightarrow$ scatter''}, which means \textit{``if the values in the data column are numerical, 
then the data column can be represented by scatter''}.}

\textbf{Scatter Plots.} The first two rules regarding scatter are straightforward. 
As suggested in~\cite{showme2007mackinlay}, scatter should be recommended when two columns of data are both numerical.
It is also a common practice to use scatter to identify anomalies, which verifies our rule~\textit{``if outliers are detected in a column with the 1.5IQR rule, then the column can be visualized by scatter''.}
This rule also matches conclusions in an empirical study~\cite{effectiveness2019saket}, as will be further discussed in Section~\ref{sec:comparison}.

\textbf{Box Plots.} Then we present two rules related to box plots. They both reflect how values in a data column are distributed.
These rules are out of expectation but make sense.
It has been seldom proposed that the values represented by box plots are not evenly distributed.
However, it is a reasonable rule. 
Box plots are often used to help discover and present the characteristic of data distribution.
In most cases, we care more about the distribution of data when it is not evenly distributed.
Thus, as our examples in Fig.~\ref{fig:gallery}(c)-(d) demonstrate, it is suitable to recommend box plot to users when the data is not evenly distributed.
The other rule \textit{``the entropy of values is large $\rightarrow$ box''} indicates that 
In other words, 
the values represented by box plots are often disordered.
This rule is also reasonable as it is not quite meaningful to inspect values' distribution if these values are too concentrated.

\textbf{Line Charts.} The next three rules show conditions to recommend  a line chart as the visualization type for a data column.
Since line charts are often applied on time series data, it is natural that the column representing time is monotonic and the data type of other columns is decimal.
Also, according to~\cite{effectiveness2019saket}, line is the best choice to represent the correlation between two series of data.
Thus, the column which is supposed to be visualized by line charts should not be the only column in the dataset.
These rules are also applied on both our generated visualizations in Fig.~\ref{fig:gallery}(e)-(f).

\textbf{Bar Charts.} The last two rules for bar charts are also intuitive.
When we use bar charts, an axis often represents a series of categorical data~\cite{showme2007mackinlay} like countries or brands. 
It is quite common that these categorical values are unique, for example, the values on x-axes of our generated visualizations in Fig.~\ref{fig:gallery}(g)-(h).
Furthermore, as the values shown in blue bars and red bars in Fig.~\ref{fig:gallery}(g) suggest, the values represented by bars can often be unsorted.

These rules and visualizations show that our method can derive meaningful rules in a data-driven manner and then recommend appropriate visualizations.
Furthermore, the unexpected but reasonable rules for box plots demonstrate that our method has the potential to help researchers identify implicit rules which map data features to appropriate visualization design choices.

\input{source/user_study}

\subsubsection{Comparison between Generated and Empirical Rules}\label{sec:comparison}

\begin{figure}[ht!]
    \centering
    \includegraphics[width=0.95\linewidth]{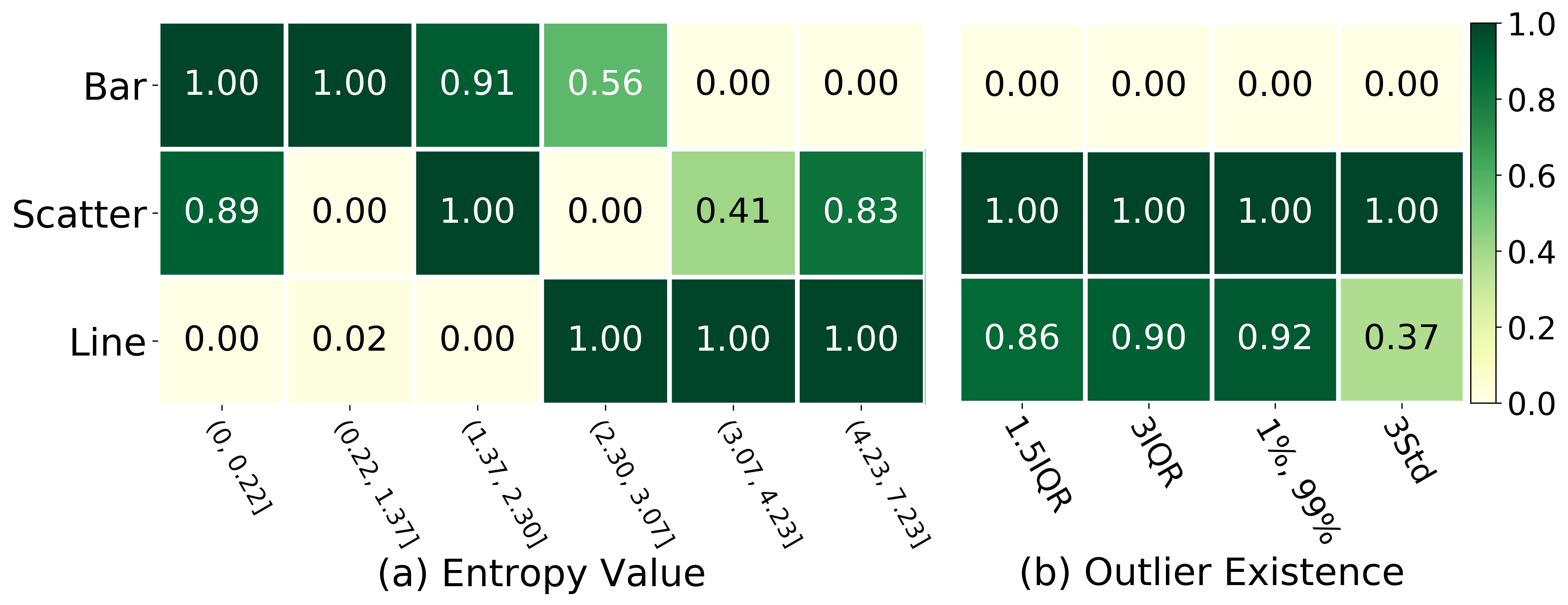}
    \vspace{-1em}
    \caption{The two heatmaps show the normalized scores of rules about (a) entropy value and (b) outlier existence. Both heatmaps share the same y-axis and the same color scale.
    A higher score and a darker color mean the visualization type is more preferred. The values on the horizontal axis of (a) denote intervals of entropy values while the values on the horizontal axis of (b) represent different criteria of outlier detection.}
    \label{fig:comparison}
    \vspace{-1em}
\end{figure}

One advantage of our method is that our recommendation results 
are explainable through
a set of extracted rules and can be easily understood by users.
So it is necessary to verify whether the extracted rules by our approach can align well with other empirical studies. 
According to our survey, Saket et al.~\cite{effectiveness2019saket} have done a crowdsourced experiment to evaluate the effectiveness of five basic visualizations (i.e., table, line, bar, scatter, and pie) that has a large overlap of visualization types with our study.
Therefore, we choose to compare the generated rules by our method with the empirical rules from this study.
Three out of five general guidelines of this study~\cite{effectiveness2019saket} are related to our research:
\begin{compactenum}
    \item \textit{Bar charts are suitable for identifying clusters;}
    \item \textit{Line charts are appropriate when finding correlations;}
    \item \textit{Scatter plots should be used to find anomalies.}
\end{compactenum}
We compare the rules above with our generated rules as follows:

\textbf{Rule 1.}
~The first rule above suggests that bar is the most effective visualization type to find clusters.
Among all the data features defined in our approach,
entropy is the feature that is more related to clusters.
A larger entropy indicates that the data in the column is more disordered, meaning that the data probably does not have obvious clusters.
Using the method described in Section~\ref{sec:inference}, we generated rules which map the value of entropy to three visualization types~(i.e., bar, line, scatter). 
The reason why only three visualization types are discussed is that the other three types of visualizations (i.e., box, heatmap, histogram) are not discussed in~\cite{effectiveness2019saket}.
The normalized scores of those rules are shown in Fig.~\ref{fig:comparison}(a).
From the figure, we can notice that, among the three visualization types, bar is the best choice when entropy is low.
Then along with the growth of entropy, scatter becomes the best choice and finally, line becomes the top choice among these three visualization types.
Since a smaller entropy indicates more obvious clusters, our generated rule actually suggests that bar should be recommended when obvious clusters exist, which aligns well with \textit{Rule 1} of the prior study.

\textbf{Rule 2.}~The second rule has been discussed in Section~\ref{sec:case_study}. 
This rule implies that there should be at least two data columns
in the dataset when recommending a line chart, as correlation is defined for two variables.
Such an idea is reflected in our extracted rule in Fig.~\ref{fig:gallery}, i.e., \textit{``the column is not the only column in the dataset $\rightarrow$ line''}.

\textbf{Rule 3.}
~According to our learned rules shown in Fig.~\ref{fig:comparison}(b), when there are outliers, scatter is always the most preferred visualization type among three common visualization types~(i.e., bar, line and scatter), which is indicated by the highest normalized scores. Thus, the generated rules by \revision{{\techNameNoSpace}} also perfectly align with \textit{Rule 3}.

In summary, the above observations confirm that {\techNameNoSpace} can extract reasonable rules that match empirical visualization guidelines.

%% file: source/user_study.tex
\subsubsection{Expert Interviews} \label{sec:user_study}
To further verify the effectiveness of recommended design choices and the correctness of generated rules, we conducted expert interviews. 

\textbf{Tasks.} In our expert interviews, we designed three tasks to evaluate the quality of rules and recommended visualizations.
\begin{compactenum}
    \item We presented the experts with the top five rules of each visualization type and  asked them to give each rule a score ranging from 1~(the least reasonable) to 5~(the most reasonable).
    \item The experts are presented with 30 datasets and the corresponding top two recommended visualizations by our approach. They were asked to give each recommended visualization a score ranging from 1~(the least reasonable) to 5~(the most reasonable). Here only the visualization type and the arrangement of x- and y-axes were required to be considered since other design choices~(e.g., the position of the legend, color usage) are not in our scope. 
    \item We presented 30 datasets to experts and let them select the top two visualization types from all six types supported by our approach to visualize the current dataset. 
    \end{compactenum}
Among three tasks, Task 1 was designed to verify the correctness of individual rules on visualization types and Task 2 aimed to evaluate the overall quality of recommended visualizations including the visualization type and axes where data columns are encoded.
Task 3 was a supplementary task to collect experts' preferred design choices. In all tasks, the sequences of visualization types or datasets were randomized.

\textbf{Datasets.} 
In our expert interviews, we presented 30 rules and 30 datasets to experts in total.
In the selection of rules presented in Task 1, we first pruned rules of the same $f_i$ and only kept the strongest rule of each feature entity.
To be more specific, after the pruning, we only kept rules with the top five highest scores in $\{f_i \rightarrow v_n | v_n \in \textrm{all visualization types}\}$ for a specific $f_i$.
Then, we presented the top five rules for each visualization type to participants for verification.
For Task 2 and Task 3, we randomly sampled 5 valid datasets for each visualization type. 
Here a valid dataset is defined as a dataset that can be correctly rendered by Plotly.
In the questionnaire of our interview, we followed the study by Hu~\textit{et al.}~\cite{vizml2019hu} to show a screenshot of the first 10 rows and the first 10 columns of each dataset to participants.
For each dataset, we presented the top two recommended visualizations to experts since the mean rank of correct visualization types is about 2 
according to Table~\ref{table:quantitative_evaluation}.
The recommended visualizations were 
exported as images with resolutions of $700 \times 450$.

\textbf{Participants and Procedure.}
We invited 12 researchers (4 females, $age_{mean}=25.42$, $age_{std}=1.93$) 
who have conducted research in data visualization for at least 1 year.
Due to the current COVID-19 pandemic, all expert interviews were conducted through online meetings.
The length of an expert interview was about 1 hour. 
Before starting the interview, we collected experts' consent for collecting their feedback and recording audio.
Each expert interview started with a 5-min brief introduction to our entire project.
After that, experts were asked to finish the three tasks.
For Task 2 and Task 3, to ensure that experts provide effective feedback for each question, they were only allowed to submit their answers on each dataset after 10 seconds.
After finishing all three tasks, experts were asked to provide general comments on our approach including the advantages and disadvantages.

\textbf{Feedback on the Generated Rules.} Overall, our generated rules are appreciated by experts. An expert commented that \textit{``some rules are inspiring''} and another expert said the rules are \textit{``straightforward and easy to understand''}.
The rules with the highest average scores are shown in Fig.~\ref{fig:gallery}.
\revision{All the rules with the lowest ($< 2$) average scores are shown in Table~\ref{table:rules}. 
}
\begin{table}[h!]
    \centering
    \small
    \setlength{\aboverulesep}{0.5pt}
    \setlength{\belowrulesep}{0.5pt}
    \caption{This table shows the five rules of the lowest scores. Most of the rules of low scores are related to the names of columns.}
    \begin{tabular}{p{5.3cm}|p{1.2cm}|p{1cm}}
    \toprule
    Feature & Type & \revision{Score} \\ \midrule
     Only one word is in the column name. & Heatmap & \revision{1.3333}\\
     The column name is started with an lower case. & Heatmap & \revision{1.6667}  \\
     The column name is less than 5 characters. &  Heatmap & \revision{1.6667} \\
     No upper case is in the column name. &  Heatmap & \revision{1.6667}\\ 
     A digit is in the the column name. & Scatter & \revision{1.8333}\\
     \bottomrule
    \end{tabular}
    \label{table:rules}
\end{table}

In Table~\ref{table:rules}, we conclude that most of the rules with low scores are related to the names of columns, which also matches experts' comments, for example, \textit{``it is not suitable to use names of columns to decide how the column should be visualized''} and \textit{``hard to convince the column name's rule has a relationship with heatmap''}.
After a careful inspection on these rules, we found that most of the rules related to column names are highly related to the grammar or the default setting of Plotly.
Though they are useful in the prediction of visualization types on datasets crawled from Plotly, they may not be able to convince general users.
For example, in our extracted rules, there are several rules related to Heatmap like the number of words in the column name, the number of characters in the column name and lack of uppercase in the column name.
They truly make sense when the datasets are from Plotly since lots of datasets visualized by heatmaps have columns named using few lower-case letters  such as ``x'' or ``y''.
However, it is quite doubtful if these rules can be generalized to other datasets so our participants considered them as unreasonable rules.
This type of unreasonable rules are mainly led by the characteristic of our dataset-visualization corpus
and this issue will be further discussed in \revision{Section~\ref{sec:data-driven}}.

\textbf{Feedback on Recommended Visualizations.} According to experts' feedback, our recommended visualizations are thought to be of high quality and can lower the burden of manually creating visualizations when exploring the dataset and designing visualizations. For example, an expert concludes our approach as an \textit{``automated method to visualize dataset without human intervention, which can ease the human workload''}.
In user interviews, we presented 2 recommended visualizations for each dataset and these visualizations are separately rated by experts.
To evaluate the overall reasonable level of our recommendation on one dataset, we use the higher score between two scores of recommended visualizations on a dataset as the final score.
The average final score of our recommendation is 3.7944, which 
shows that the experts think most of our recommended visualizations are reasonable.
Detailed score distribution of visualization types is also shown in Fig.~\ref{fig:average_score}.
From Fig.~\ref{fig:average_score}, we can learn that our approach performs well on bar, histogram, line and scatter but performs relatively worse on box and heatmap.
The cases with the highest scores are shown in Fig.~\ref{fig:gallery}.

\begin{figure}[h!]
    \centering
    \includegraphics[width=0.9\linewidth]{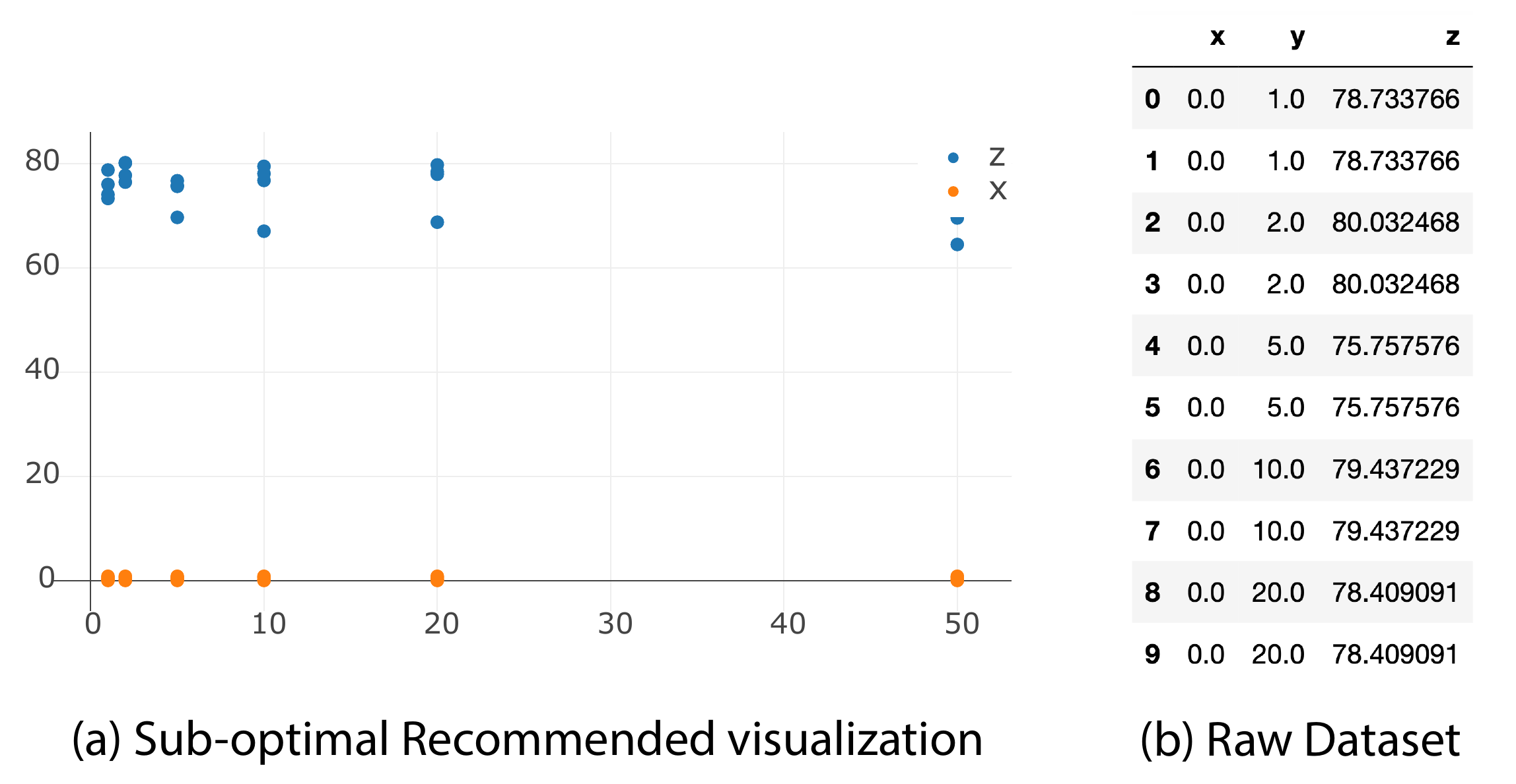}
    \vspace{-1em}
    \caption{This figure shows (a) a sub-optimal recommended visualization identified in expert interviews and (b) the first 10 rows of its raw dataset.}
    \label{fig:failure_case}
\end{figure}

\begin{figure}[h!]
\begin{minipage}[t]{0.48\linewidth}
    \includegraphics[width=\linewidth]{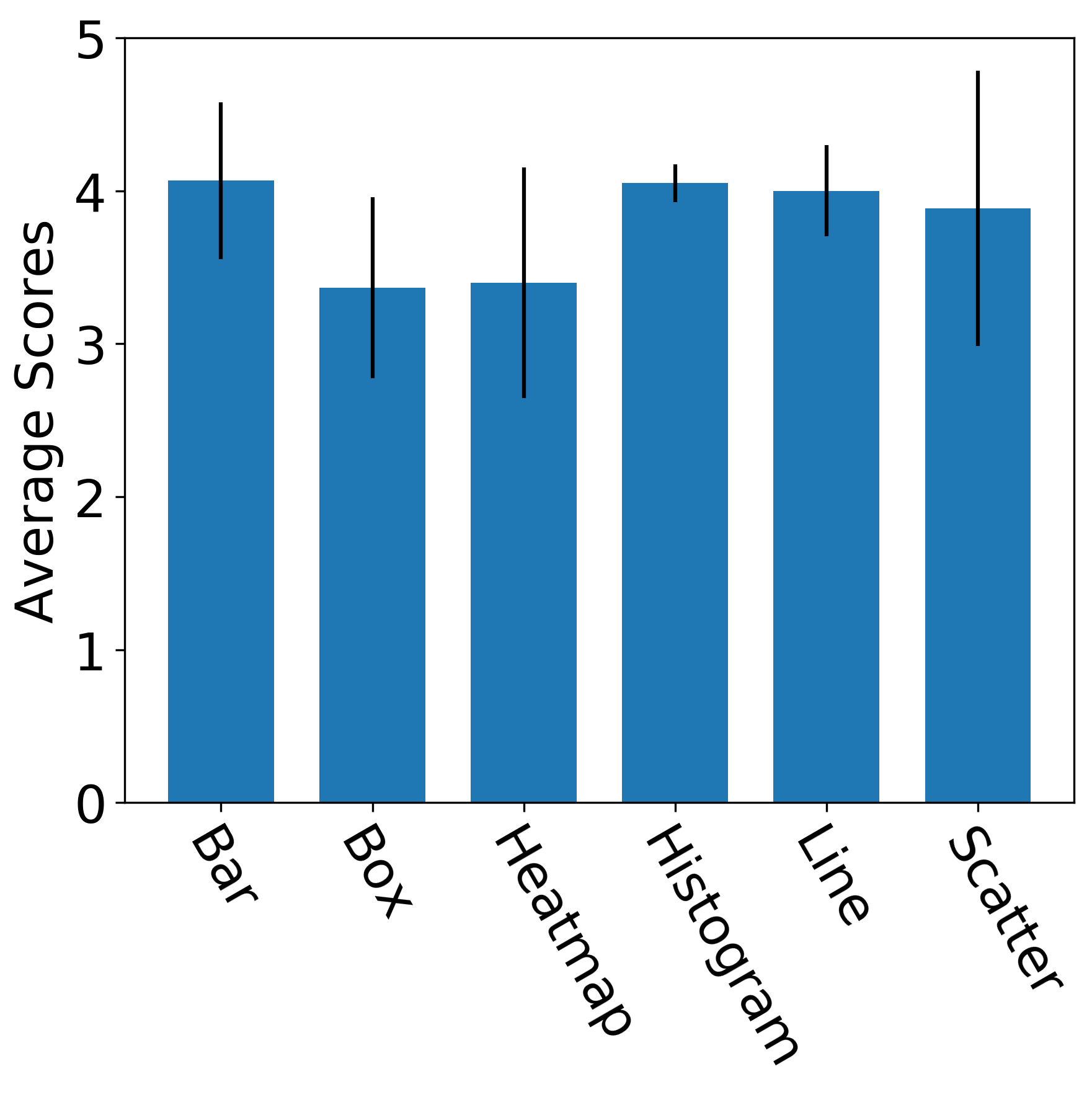}
    \vspace{-1.5em}
    \caption{Average scores of recommended visualizations. \revision{The vertical lines on the top of the bars denote the standard deviation values.}}
    \label{fig:average_score}
\end{minipage}%
    \hfill%
\begin{minipage}[t]{0.48\linewidth}
    \includegraphics[width=\linewidth]{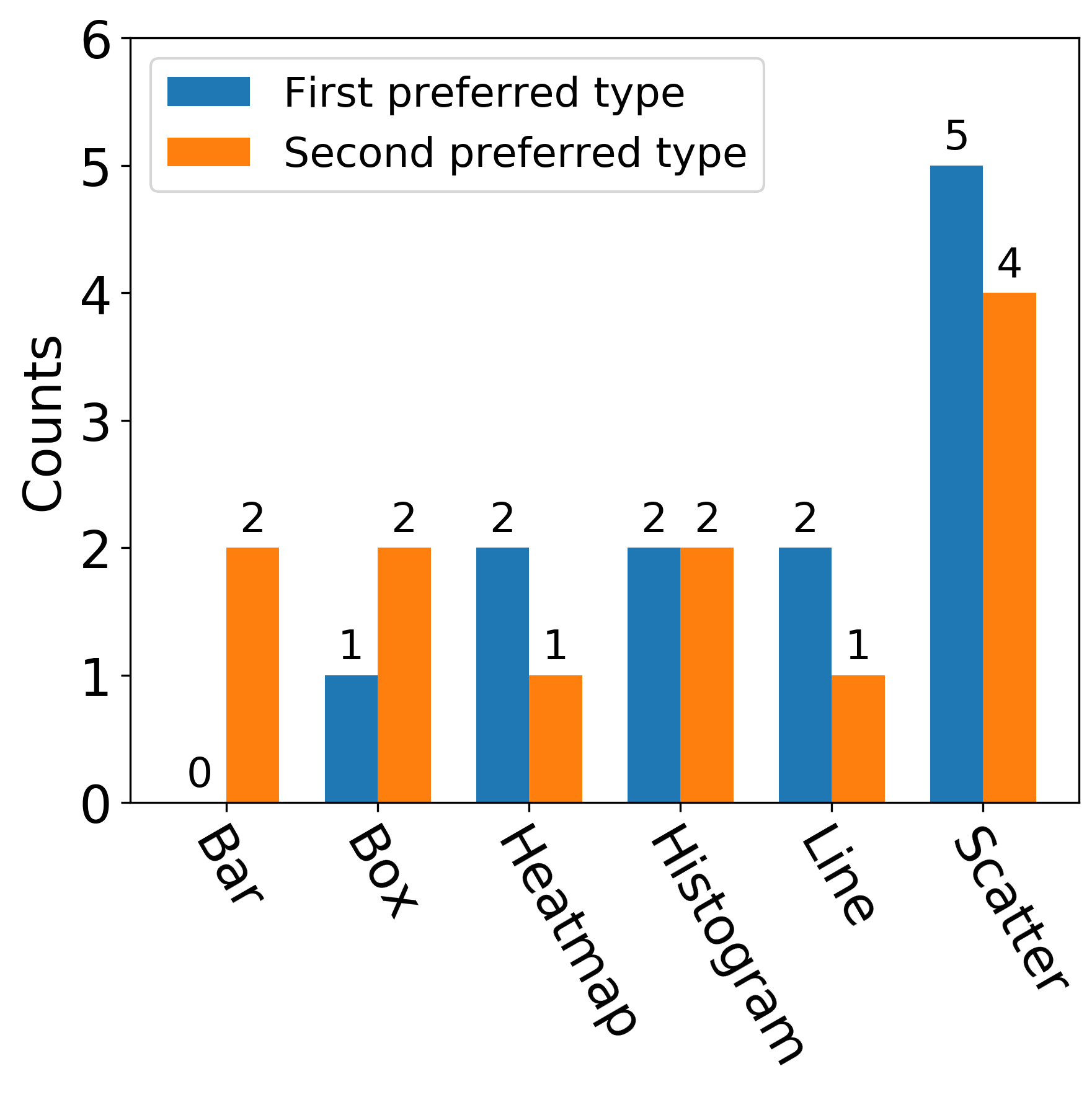}
    \vspace{-1.5em}
    \caption{Experts' preferred visualization types of the dataset in Fig.~\ref{fig:failure_case}\revision{(b)}.}
    \label{fig:preference}
\end{minipage} 
\vspace{-1em}
\end{figure}

To further understand sub-optimal recommended visualizations which are originally visualized by box plots and heatmaps, we checked these sub-optimal cases carefully and speculated that these datasets lack dominant features to determine their optimal visualization type and the visualization design choice of this dataset may depend on users' preference and analysis tasks.
Fig.~\ref{fig:failure_case} illustrates a case which is originally visualized by a heatmap and got a low score in our expert interview.
To better analyze this failure case, 
we checked experts' preferred visualization type of this dataset in Task 3 and the results are shown in Fig.~\ref{fig:preference}.
The results show that experts' opinion on the best type of visualization was quite inconsistent, which is also reflected by the entropy of selections.
The entropy values of both the best and the second best visualization types are larger than average values~($entropy^{top_1} = 1.4677$, $entropy^{top_2} = 1.6762$, $entropy^{top_1}_{mean} = 0.9868$, $entropy^{top_2}_{mean} = 1.4515$). To handle this kind of dataset, it is possible to further consider users' preferences and their specific tasks in our inference method to achieve better recommendation results, which is left as our future work.

%% file: source/discussion.tex
\section{Discussion}
In this section, we summarize the lessons we learned during the development of \techName and discuss the limitations of our approach.

\subsection{Lessons}\label{sec:lessons}

By modeling the relation between data features and effective visualizations, we have learned many lessons from building knowledge graphs for visualization recommendation.

\subsubsection{Knowledge Graph for Visualizations}
Knowledge graph aims to build a structured form of human knowledge and has drawn great research attention from both the artificial intelligence (AI) research community and the industry~\cite{surveyKG2020ji,wang2017knowledge,hogan2020knowledge}.
However, little research has been done on building knowledge graphs for modeling the visualization design knowledge and common practices in real visualization applications.
Our work is the first study along this direction. Two key steps have been crucial to the construction of knowledge graphs for visualization, i.e., \textit{KG construction} and \textit{embedding learning}. 

\textbf{Entity Construction.} Entity construction in this paper aims to create entities to delineate continuous data features, which is intrinsically required by the structure of knowledge graphs. Those categorical data features (e.g., data attribute types) can be directly regarded as entities.
But for data features with a continuous value, it is necessary to discretize them, so that they can be represented as entities
which describe specific characteristics of a data feature, as discussed in Section~\ref{sec:kg_construction}.
We employ the MDLP technique for the discretization of continuous data features, as it collectively considers both the distribution of each data feature and their overall correspondence with different visualization types.
Before using MDLP, we also explored other techniques such as KMeans for feature discretization. But KMeans considers only the distribution of each individual feature, resulting in an inferior recommendation performance than MDLP.
Thus, MDLP is finally chosen in our approach.
Other advanced discretization techniques can be further explored, which, however, is beyond the major focus of this paper. 

\textbf{Embedding Learning.} 
When developing \techNameNoSpace, we have considered three types of embedding learning techniques including TransE, TransE-adv and RotatE and finally adopted TransE-adv.
The reason why these three methods are considered is that these methods are widely recognized, efficient and intuitive.
Compared with other models like TransR~\cite{transr2015lin} and RESCAL~\cite{rescal2011nickel}, these three models have lower time and space complexity~\cite{wang2017knowledge}.
Also, their methods of modeling relations are intuitive. They model relations between entities as  translation or rotation in the space. This can help users better interpret the learned embeddings and generated rules.

\subsubsection{Data-driven Visualization Recommendation}\label{sec:data-driven}

Similar to the ML-based methods for visualization recommendation~\cite{vizml2019hu, data2vis2019dibia}, 
\techName also learns from existing visualization examples and does not require users to manually specify a list of visualization rules.
Our evaluations above have demonstrated the usefulness and effectiveness of \techNameNoSpace. It benefits from two main perspectives and can be further improved: the \textit{corpus of dataset-visualization pairs} and \textit{explainability}.
\revision{We will also further clarify its \textit{pros and cons} compared with other
visualization recommendation approaches in this section.
}

\textbf{Corpus.} 
\techName is a data-driven approach for visualization recommendation, which intrinsically depends on dataset-visualization pairs. 
\techName is built on the VizML corpus~\cite{vizml2019hu} to construct knowledge graphs for visualizations.
In the past few years, there has been an increasing number of open-sourced visualization corpora~\cite{vizml2019hu, poco2017reverse, deng2020visimages, chen2021vis30k, yuan2021deep, chen2020composition, chen2020infographic}. 
These visualization corpora have made data-driven visualization recommendations or automated visualization designs feasible.
On the other hand, we also noticed that these corpora suffer from some drawbacks, including the dependence on specific visualization grammar and limited visualization types, for example, the VizML corpus~\cite{vizml2019hu} and the Vega-Lite corpus~\cite{poco2017reverse}.
More high-quality corpora with diverse visualization examples will further benefit the data-driven approaches for visualization recommendation.

\textbf{Explainability.}
The visualization recommendations by \techName can be explained by a set of rules, which is highly appreciated by users as shown in our user study. These rules explicitly inform users of common visualization practices and are helpful for users without a background of visualization to know why specific visualizations are recommended for a given dataset.
\revision{Meanwhile, we also noticed two important issue within those rules that may 
hinder them from quickly understanding the rules: \textit{the feature complexity} and \textit{the number of rule conditions}.
First, some data features are not that straightforward to general users who
may have no knowledge of some data features.
For example, our data features include the 
moments\footnote{\href{https://en.wikipedia.org/wiki/Moment\_(mathematics)}{https://en.wikipedia.org/wiki/Moment\_(mathematics)}} of the distribution of a data column. 
The moments are used to describe the shape of data distribution
and are widely applied in visualization recommendation approaches~\cite{vizml2019hu, demiralp2017foresight}.
However, the detailed statistical meaning of moments may be unclear to some users and the related rules may confuse them.
}
Thus, it may be necessary to strike a balance between expressiveness and intuitiveness when choosing data features. 
\revisioncr{ Also, how to present the rules to users in an automatic and more understandable way is worth further exploration.}

\revision{
Second, the number of conditions in one rule can also affect the intuitiveness and expressiveness of rules.
A rule with multiple conditions is more expressive but less understandable for general users.
Meanwhile, a rule with only one condition is more straightforward but not sufficient to capture the characteristics of data columns, negatively affecting the visualization recommendation. 
To address this issue, we have proposed the approach described in Section~\ref{sec:inference} to consider multiple conditions comprehensively by aggregating all the valid one-condition rules and further recommend visualization design choices. 
With this approach, we are able to present users with understandable rules with one condition and also guarantee an effective visualization recommendation.}

\textbf{Pros and Cons.} \revision{Compared with existing rule-based and ML-based visualization recommendation methods, \techName has its pros and cons.}
\revision{
Compared with the rule-based methods~(e.g., APT~\cite{apt1986machinlay}, Show Me~\cite{showme2007mackinlay}), \techName is also explainable, but has better extendability,
since \techName can derive rules automatically from an existing corpus of dataset-visualization pairs.
Meanwhile, the corresponding limitation of \techName is that the quality and coverage of the rules depend on the dataset-visualization corpus, as discussed previously.
}
\revision{
Similar to other ML-based methods~(e.g., Data2Vis~\cite{data2vis2019dibia}, VizML~\cite{vizml2019hu}), \techName can also achieve data-driven and real-time~(average time is 0.07s per dataset) visualization recommendation for a new dataset. Furthermore, it guarantees the explainability of the recommendation by utilizing the derived rules.
However, the performance of \techName is slightly lower than the neural network used in VizML~\cite{vizml2019hu}~(MR of the visualization type is 1.7755), which can be further improved.
The major reason may be that \techName does not delineate non-linear relations between data features and design choices as well as neural networks. 
}

\subsection{Limitations}\label{sec:limitation}

Our evaluations in Section~\ref{sec-evaluation} have demonstrated the effectiveness of {\techNameNoSpace}. However, it is not without limitations.

\textbf{Limited Visualization Design Choices.}
For visualization design choices, we mainly consider the visualization types and the arrangement of x-/y-axes. 
However, there are more advanced visualization design choices when making charts, for example, the ratio of the chart and the usage of color.
\revision{The} major reason why we do not consider these design choices is as below.
Some advanced design choices such as the usage of colors require more professional knowledge~\cite{yuan2021infocolorizer}, which may not be able to be learned in the current corpus.
For example, according to our observation, most of the visualizations 
in our corpus do not define colors explicitly and 
only use the default color scale of Plotly.
\revision{
To address this issue, some corpora containing other design choices can be used to augment the current KG.
The KG used in \techName can be easily extended by connecting the entities of \textit{data features}, \textit{design choices} and \textit{data columns}. 
But when more data features and design choices are introduced, the number of triplets in the KG will grow linearly and a longer training time should be expected.
}

\textbf{Evaluations.} 
Though we have conducted extensive evaluations for our methods, one limitation of our evaluations is that we conduct interviews with only 12 experts.
However, it is not very feasible to conduct our interviews at a larger scale, for example, using Amazon MTurk.
The main reason is that participants are asked to verify the correctness of our generated rules, which
requires that the participants should have enough knowledge in visualization design to judge the rules.
Thus, we believe that this task is not suitable for MTurk workers, as they are not necessarily visualization experts.

\section{Conclusion and Future Work}
\label{sec:future_work}
We propose \techNameNoSpace, a knowledge graph~(KG)-based approach to achieve automated and explainable visualization recommendation, which can
effectively capture the visualization knowledge from dataset-visualization pairs and provide users with meaningful rules
to help them understand the visualization recommendation results.
{\techNameNoSpace} consists of four key
modules:
feature extraction, 
KG construction, embedding learning and embedding-based inference.
First, expressive data features are extracted for dataset-visualization pairs.
Then, we
build a KG with three types of nodes, i.e., data features, data columns and visualization design choices.
Further, TransE-adv is employed to learn the embeddings of entities and relations.
Based on the learned embeddings, 
we finally propose an inference method that can recommend visualizations for new datasets and generate visualization rules for explaining the recommendation results.
We conducted extensive evaluations to demonstrate the effectiveness of {\techNameNoSpace}, including case studies, expert interviews and comparisons between the generated rules by our approach and empirical rules from a prior study.

In future work,
we plan to explore how to incorporate cross-column features in our KG without increasing the need for computational power.
Also, we would like to further investigate how to incorporate different user requirements and preferences to achieve personalized visualization recommendations to different users.
Furthermore, it will be interesting to extend the proposed KG-based visualization recommendation approach to other types of visualizations such as infographics.